\documentclass{article}
\usepackage{fortschritte}
\usepackage{amsmath}
\usepackage{hyperref}

\newdimen\tableauside\tableauside=1ex   
\newdimen\tableaurule\tableaurule=.32pt   
\newdimen\tableaustep
\def\phantomhrule#1{\hbox{\vbox to0pt{\hrule height\tableaurule width#1\vss}}}
\def\phantomvrule#1{\vbox{\hbox to0pt{\vrule width\tableaurule height#1\hss}}}
\def\sqr{\vbox{%
  \phantomhrule\tableaustep
  \hbox{\phantomvrule\tableaustep\kern\tableaustep\phantomvrule\tableaustep}%
  \hbox{\vbox{\phantomhrule\tableauside}\kern-\tableaurule}}}
\def\squares#1{\hbox{\count0=#1\noindent\loop\sqr
  \advance\count0 by-1 \ifnum\count0>0\repeat}}
\def\tableau#1{\vcenter{\offinterlineskip
  \tableaustep=\tableauside\advance\tableaustep by-\tableaurule
  \kern\normallineskip\hbox
    {\kern\normallineskip\vbox
      {\gettableau#1 0 }%
     \kern\normallineskip\kern\tableaurule}%
  \kern\normallineskip\kern\tableaurule}}
\def\gettableau#1 {\ifnum#1=0\let\next=\null\else
  \squares{#1}\let\next=\gettableau\fi\next}

\newcommand{\eqn}[1]{(\ref{#1})}  

\begin{document}
\def\titleline{Towards complete string effective actions beyond leading order}
\def\email_speaker{{\tt kasper.peeters@aei.mpg.de}}
\def\authors{Kasper Peeters\1ad\hskip-.2em\sp\hskip-.2em, Pierre Vanhove\2ad and Anders Westerberg\3ad}
\def\addresses{
\1ad MPI/AEI f{\"u}r Gravitationsphysik, Am M{\"u}hlenberg 1, 14476 Golm, Germany\\
\2ad CEA/DSM/SPhT, URA au CNRS, CEA/Saclay, F-91191 Gif-sur-Yvette, France\\
\3ad Department of Physics, Karlstad University, S-65188 Karlstad, Sweden
}
\def\abstracttext{
We review the current knowledge of higher-derivative terms in string
effective actions, the various approaches that have been used to
obtain them and their applications.
}
\large
\makefront
\section{Higher-derivative string effective actions}

Many aspects of low-energy string dynamics can be captured in terms of a 
Wilsonian effective gravity field theory for the massless modes. Although
it is often sufficient to consider only the lowest-order supergravity actions, 
there are qualitative string predictions for which knowledge about the
subleading terms is required. Despite the fact that the first results about 
the structure of higher-derivative string effective actions are almost twenty 
years old, complete, supersymmetric invariants are still lacking, even at the 
first sub-leading order in~$\alpha'$ beyond the supergravity level.

Four main techniques have been employed to derive supersymmetric
higher-derivative string effective actions. One is to simply try to
construct the most general supersymmetric invariant containing higher
derivatives. Such an approach was pursued in the early
days~\cite{dero3}, but it is extremely cumbersome (even on a computer)
due to the enormous number of terms and the problem of dealing with
side relations such as Bianchi and Ricci identities, as well as
partial integration. A second method employs the relation between
string background field equations of motion and conformal invariance
on the world-sheet. This method is very useful for the NS-NS sector of
string theory, for which background field couplings are under control
in the RNS formulation of the string, but it becomes much more
complicated for R-R background fields. The latter can at present only
be treated in the Berkovits formalism; the state of the art on such
calculations can be found in~\cite{Berkovits:2001ue}.  Thirdly, the
standard supergravity technique of solving superspace Bianchi
identities is being pursued, with the higher-derivative terms arising
from relaxed torsion constraints~\cite{howe7,Cederwall:2000ye}. While
such an approach is potentially very useful for the construction of
existence or uniqueness proofs concerning higher-derivative
invariants, obtaining explicit component-space expressions in this way
is very difficult.  At present, the only method which has shown
promise to be powerful enough to determine, in practise, the
\emph{entire} effective action in component form at sub-leading order
is the construction directly from string scattering amplitudes. In the
present letter we review the current status of this approach and
comment on some related problems and often-raised questions.

\section{Killing spinors and supersymmetry-preserving solutions}

One reason to study higher-derivative corrections to supergravity
actions is that a knowledge of them would allow us to study
corrections to supergravity solutions, e.g.~black holes or D-branes.
Such corrections are important because they influence entropy counting
(for an analysis in four dimensions see \cite{LopesCardoso:1999ur}) or
predictions of the string/gauge theory correspondence (see
e.g.~\cite{Frolov:2001xr} for some initial steps in this
direction). Many of these solutions contain non-trivial gauge-field
configurations. A full analysis is at present seriously hampered by
our incomplete understanding of the higher-derivative bosonic terms in
the effective action.

Interesting solutions are those which preserve at least some fraction
of supersymmetry, i.e.~backgrounds in which the variations of the
fermions vanish. For the gravitino in M-theory this takes, at
classical level, the form
\begin{equation}
\label{e:SuperCov}
\delta_\epsilon\psi_\mu =   \Big(\partial_\mu
+\tfrac{1}{4}\omega_{\mu\nu_1\nu_2} \, \Gamma^{\nu_1\nu_2}+
T_\mu{}^{\nu_1\cdots\nu_4}  F_{\nu_1\ldots\nu_4}\Big) \epsilon = 0\, ,
\end{equation}
where $F={\rm d}C$ and
$T_\mu{}^{\nu_1\cdots\nu_4}=(\Gamma_\mu{}^{\nu_1\cdots \nu_4}-8
\delta_\mu{}^{\nu_1} \Gamma^{\nu_2\cdots \nu_4})/288$. This equation
gives rise to a set of integrability conditions
$[\mathcal{D}_\mu,\mathcal{D}_\nu]\epsilon=0$. These equations,
together with the equation of motion and Bianchi identity for the
four-form, imply the equation of motion for the metric. The existence
of a Killing spinor satisfying~\eqn{e:SuperCov} imposes severe
restrictions on the holonomy group associated with the vacuum
solution, and restricts it to a subgroup of the Lorentz group.

Higher-derivative corrections to the action do, however,
modify~\eqn{e:SuperCov}, because superinvariance at higher order means
that there are corrections to the supersymmetry transformation rules:
invariance of the action means that
\begin{equation}
\Big(\delta_0 + \sum_n (l_{\rm P})^n \delta_n\Big)\,\Big( S_0 + \sum_{n}
(l_{\rm P})^n  S_n\Big)=0\, . 
\end{equation}
More explicitly, the Killing spinor equation receives corrections of
the form
\begin{equation}
\delta\psi_\mu = 
  \Big(\nabla_\mu + T_\mu{}^{\nu_1\cdots\nu_4}
  F_{\nu_1\ldots\nu_4}\Big) \epsilon + (l_{\rm P})^6 
  \big( DR^3 \epsilon \big)_\mu + (l_{\rm P})^6 \big( \ldots \big)_\mu\, .
\end{equation}
In certain situations, such as compactifications on large
eight-manifolds, it is consistent to ignore all the $l_{\rm P}$
correction terms. This can be seen from a simple scaling argument.
Under a scaling of the eight-manifold $g_{(8)}\rightarrow t g_{(8)}$
the first two terms scale as~$t^{-3/2}$. The third term, which is the
$C\wedge R^4$-induced correction to the supersymmetry transformation
rules~\cite{Peeters:2000qj}, instead scales as~$t^{-3}$. Therefore it
is perfectly consistent, for large~$t$, to use only the lowest order
Killing spinor equations~\eqn{e:SuperCov}, as was done in
e.g.~\cite{Becker:1996gj}. This is true despite the fact that the
four-form equation of motion \emph{does} receive a correction from the
$C\wedge t_8 R^4$ term in the action,
\begin{equation}
\label{e:DHeom}
{\rm d}{*}F_4 = \tfrac{1}{2}F_4\wedge F_4 + (l_{\rm P})^6 \,t_8 R^4 + (l_{\rm P})^6 \big(\ldots\big) \, ,
\end{equation}
where the suppressed terms involve fields other than the graviton (and
are at present completely unknown).  Here the terms which are listed
explicitly in~\eqn{e:DHeom} are all scale invariant. Any other terms,
for instance those arising from an~$R^3 F^2$ term in the effective
action, scale with some negative power of~$t$ and will therefore be
suppressed in the large-volume limit. However, in general
applications, these correction terms cannot be scaled away.  Several
corrections to the supersymmetry transformation rules have been
computed by us some time ago~\cite{Peeters:2000qj}.  Incorporating the
effects of higher-derivative terms and their influence on the holonomy
structure group is still an open problem.

Given that the left-hand side of the equation of motion~\eqn{e:DHeom}
is closed, it is natural to write it as
\begin{equation}
{\rm d} F_7 = \tfrac{1}{2} F_4 \wedge F_4 + (l_{\rm P})^6 \,t_8  R^4 
\end{equation}
(now meant to be read as an exact equation), supplemented by the
duality relation
\begin{equation}
F_7 = {*}F_4 + (l_{\rm P})^6 (\ldots)\, .
\end{equation}
This is the point of view taken in~\cite{Howe:2003cy}, where the
physically non-trivial deformations of this duality equation (or
rather its superspace version) were analysed. In this form, an
interesting parallel to the type-IIB theory
arises~\cite{Peeters:2003pv}. Instead of having two gauge fields
related by a duality condition, one there has a single gauge field with
a self-duality condition. The analogue of the deformation of the
duality relation between~$F_7$ and~$F_4$ now becomes a deformation of
the self-duality condition of $F_5$ (or rather the composite field
strength $\tilde F_5 = {\rm d}C_4 + \tfrac{1}{2} B_2\wedge F_3 -
\tfrac{1}{2} H_3 \wedge C_2$ with $H_3={\rm d}B_2$ and $F_3={\rm d}
C_2$),
\begin{equation}
 \tilde{F}_5 + (\alpha')^3 \frac{\delta S^{(3)}}{\delta\tilde F_5^+} =
 {*}\left[ \tilde{F}_5 + (\alpha')^3 \frac{\delta S^{(3)}}{\delta\tilde F_5^+} 
   \right] \,.
\end{equation}
Some of these corrections have recently been computed by two
of us~\cite{Peeters:2003pv}.

\section{Actions from string amplitudes}
\label{s:amplitudes}

As stressed before, our approach is to derive effective actions
directly from string theory amplitudes. At genus zero and one, the
amplitudes with four external gravitons have been known for quite some
time; these are relatively easy to compute and lead to the well-known
fourth-order action
\begin{equation}
\label{e:boringt8t8}
S = \int\!{\rm d}^{10}x\,\sqrt{-g} t_8 t_8 R^4\, .
\end{equation}
This term is universal for all string theories. Beyond the
four-point NS-NS sector, which also includes the two-form field,
the situation quickly becomes much more complicated.

There are several reasons why such amplitudes are hard to compute. One
of them is that the vertex operators for R-R gauge fields involve spin
fields, the presence of which makes the worldsheet fermionic
correlators difficult to evaluate. Fortunately, generic expressions
have been derived~\cite{Atick:1987rs} which completely resolve this
problem and which avoid any explicit operator product
expansions.

A second and more serious problem concerns the integration over the
odd supermoduli, or more simply put, the integration over the
inequivalent sectors of the world-sheet gravitino field. Consider a
generic correlator, say at genus one in the odd/odd spin-structure
sector, expanded in powers of the gauge-fixed world-sheet gravitinos
$\chi_-$ and $\tilde\chi_+$:
\begin{multline}
\label{e:generic_corr}
\Big\langle V_1(z_1)\cdots V_n(z_n)\Big\rangle = 
\int\! {\cal D}\chi{\cal D}\bar\chi
{\cal D}X {\cal D}\Psi\, \Big[V_1(z_1)\cdots V_n(z_n)\Big]\, \exp\big({-S[X,\Psi]}\big)\\[1ex]
\times e^{\phi+\tilde\phi}\,\bigg(
1 - \frac{1}{2\pi\alpha'}\int\!{\rm d}^2z\, \tilde\chi_+\Psi\partial X(z)
  - \frac{1}{2\pi\alpha'}\int\!{\rm d}^2z\, \chi_- \tilde \Psi \bar\partial X(z)
  + \frac{1}{4\pi\alpha'}\int\!{\rm d}^2z\, \tilde \chi_+\chi_-\Psi\tilde\Psi(z)\\[1ex]
  + \frac{1}{(2\pi\alpha')^2}\int\!{\rm d}^2w\!\! \int\!{\rm d}^2z\, 
     \tilde\chi_+\Psi\partial X(w) \chi_- \tilde \Psi \bar\partial X(z)
\bigg)\, .
\end{multline}
The Grassmann integrals select the terms with one~$\chi_-$ and
one~$\tilde\chi_+$. One usually therefore only considers the last two
terms in this expression, which have the structure of so-called
``picture changing operators''. However, this is only a correct
procedure if there are no world-sheet gravitino modes in the vertex
operators themselves, which is unfortunately \emph{not}
true~\cite{Peeters:2001ub} (see also the early
work~\cite{D'Hoker:1987bh}). For e.g.~the NS-NS two-form vertex
operator, these additional terms are given by
\begin{multline}
\label{e:V00B}
V^{(0,0)}_B = V^{(0,0)}_B\Big|_{\text{standard}} -\frac{1}{6}\int\!{\rm d}^2z\, H_{\mu \nu\rho}\Big(
\Psi^\mu \Psi^\nu\Psi^\rho\tilde\chi_+ - \tilde\Psi^\mu \tilde\Psi^\nu
\tilde\Psi^\rho\chi_- 
\Big)\,e^{ik_\rho X^\rho}\, .
\end{multline}
Simple examples exist in which these gravitino terms really
matter~\cite{Peeters:2001ub}, and this issue becomes more complicated
for increasing genus.

Once the correlators have been computed, one is still left with
integrals over the modular parameters and vertex operator insertion
points. Despite the loop-by-loop finiteness claim of string theory,
these modular integrals are actually divergent for external momenta in
the physical regime. This is a well-known problem, present already for
the four-graviton amplitude at tree level~\cite{D'Hoker:1995yr} and
typically circumvented by performing an analytic continuation in the
Mandelstam variables, computing the integral in terms of standard
functions, and then analytically continuing back. Unfortunately, this
procedure in general requires that the integral is cut up in various
pieces, which each have to be analytically continued in a different
way~\cite{D'Hoker:1995yr}. This entire procedure makes it very hard to
construct a systematic expansion in $\alpha'$ (let alone to do these
integrals numerically). The origin of this problem lies in the
Euclidean formulation of string perturbation theory~\cite{lorentzian}.

Having resolved several of these problems
in~\cite{Peeters:2001ub,Peeters:2003pv}, two of us have recently been
able to perform the next step in the completion of the effective
action~\eqn{e:boringt8t8} with other bosonic fields. To give a flavour of the
form into which these results can be cast\footnote{It should be
mentioned that even when the relevant amplitudes have been calculated,
the step of converting amplitudes to effective-action terms may be
quite complicated. This is, in particular, true for the R-R bilinear
higher-derivative terms to be discussed below, owing to the need to be
able to recognise terms proportional to the lowest-order equations of
motion.}, we show here the form of the genus-one contribution to~$W^2
(DF_{(5)}^+)^2$ terms in the effective action~\cite{Peeters:2003pv}:
\begin{multline}
\label{e:finalresult}
S^{W^2 (DF_{(5)}^+)^2}_{\text{IIB}} = \int\!{\rm d}^{10}x \,\sqrt{-g}\;
\bigg( 
  (16+\lambda)\, W^2\Big|_{\widetilde{\tableau{2 2 2 2}}} 
- 4(16-\lambda)\, W^2\Big|_{\widetilde{\tableau{2 2 2}}}
+192\, W^2\Big|_{\widetilde{\tableau{3 1 1 1}}}\\[1ex]
+\tfrac{16}{15}(16+\lambda)\, W^2\Big|_{\widetilde{\tableau{2 2}}_A}
+\tfrac{32}{3}\,   W^2\Big|_{\widetilde{\tableau{4}}}
+\tfrac{1}{21}(16+\lambda)\, W^2\Big|_{\widetilde{\tableau{2}}}\bigg)\,
(DF_{(5)}^+\Big|_{\widetilde{\tableau{2 1 1 1 1}}^+})^2\,.
\end{multline}
Here $\lambda$ corresponds to a one-parameter ambiguity in mapping the
string amplitudes to terms in the effective action.\footnote{There are
always ambiguities when translating on-shell string amplitudes to an
effective action. Several terms with~$n$ powers of the fields may lead
to vanishing on-shell $n$-point amplitudes, and in addition there is
the freedom of field redefinition, which can be used to change the
higher-order action by any term proportional to the lower-order
field equations.} This parameter is fixed by linearised supersymmetry
to the value $\lambda=-16$, leaving only three non-vanishing
contributions in~\eqn{e:finalresult}.

\section{Superspace and superspace constraints}

As discussed at length in~\cite{Peeters:2000qj}, there is a close
connection between supersymmetry transformation rules and superspace
torsion constraints: the algebra satisfied by the supersymmetry
variations is isomorphic to that of the supercovariant derivatives,
$\{D_a,D_b\}={T_{ab}}^r D_r$.  This link provides a way to obtain
string-/M-theory corrections to the superspace geometries of ten- and
eleven-dimensional supergravity theories through the construction of
the relevant higher-derivative supersymmetry invariants.  Such
corrections are, for instance, of central importance for the study of
higher-derivative corrections to kappa-symmetric world-volume actions
of D-branes and M-branes.

A crucial step in this programme to obtain corrections to the torsion
constraints from the component formalism was made
in~\cite{Peeters:2000qj}. This involved the use of string input to
cast the $d=10$, $N=1$ supersymmetric completion (originally derived
in~\cite{dero3}) of the term
\begin{equation}
\label{e:IX}
I_X = t_8t_8R^4 + \tfrac{1}{2}\varepsilon_{10}B t_{8} R^4 
\end{equation}
in a compact~``$t_8$'' form, together with the associated
modifications of the supersymmetry transformations of the basic
fields.  These were subsequently lifted to eleven dimensions. It was
found, however, that incorporating the~$I_X$ superinvariant does not
lead to any modifications of the superspace geometry. Using
cohomological methods in superspace~\cite{Cederwall:2001dx}, it has
subsequently been shown~\cite{Howe:2003cy} that any modification of
eleven-dimensional supergravity that does not involve a non-vanishing
lowest-dimensional component of the four-form superfield, is trivial.
Hence, in a component-space approach, we expect that the inclusion
(and supersymmetrisation) of higher-derivative gauge-field terms is
necessary in order to get explicit results for the corrections to the
superspace geometry. It is also still possible that such corrections
arise from the supersymmetrisation of the term
 \begin{equation}
\label{e:IZ}
I_Z = -\varepsilon_{10}\varepsilon_{10}R^4 + 4 \varepsilon_{10}Bt_{8} R^4 \,,
\end{equation}
appearing together with $I_X$ in the type-II effective actions.  
In~\cite{Peeters:2001ub} some partial results in this direction were
presented. More specifically, it was shown how a careful treatment of
left/right-mixing zero-mode terms in the bosonic two-point functions
on the torus is necessary to resolve an otherwise puzzling issue
regarding the cancellation mechanism for supersymmetry variations of
the anomaly term. 

Finally, let us comment briefly on the chiral superfield of the
type-IIB theory~\cite{howe1}, which has been used to construct a
superinvariant at the linearised level. If this construction is
extended to the non-linear level a chiral measure is required
\cite{deHaro:2002vk} in order to perform the integration over the odd
coordinates. Such a chiral measure is known \emph{not} to exist as
there exists only one chiral superfield~\cite{howe1}. However, this
does not necessarily prevent the construction of an on-shell
non-linear superinvariant, as the modified torsion constraints relate
the variations of the measure (or rather the full higher-derivative
part of the action) to the lowest-order supergravity action. While
examples do exist in which a linearised supergravity action based on
chiral superfields requires the introduction of non-chiral fields at
the non-linear level (e.g.~$N=3$ conformal supergravity), all of these
involve only lowest order actions and none involve mixing through
modified torsion constraints.  In any case, the scalar superfield is
somewhat of a curiosity in the larger scheme of things, and the only
rigorous way to obtain information about non-linear terms in the
effective action at the component level is, at present, by computing
them directly from string theory.

\begingroup\raggedright\endgroup
\end{document}